\pdfoutput=1
\documentclass[12pt,a4paper]{article}

\newcommand*{\BsDsDs}{\ensuremath{\Bs \to \D^{(*)+}_\squark \D^{(*)-}_\squark}\xspace}
\newcommand*{\Dsstst}{\ensuremath{D_{\squark1}(2460)^{+}}\xspace}
\newcommand*{\Dsststz}{\ensuremath{D_{\squark0}(2317)^{+}}\xspace}
\newcommand*{\BDsD}{\ensuremath{{\Bd \to D^{(*)\pm}_{s} D_{}^{(*)\mp}}}\xspace}
\newcommand*{\DsJp}{\ensuremath{D_{\squark J}^{+}\xspace}}
\newcommand*{\Tstrut}{\rule{0pt}{2.6ex}}         % = `top' strut
\usepackage{ifthen} % for conditional statements
\newboolean{pdflatex}
\setboolean{pdflatex}{true} % False for eps figures 

\newboolean{articletitles}
\setboolean{articletitles}{true} % False removes titles in references

\newboolean{uprightparticles}
\setboolean{uprightparticles}{false} %True for upright particle symbols

\newboolean{inbibliography}
\setboolean{inbibliography}{False} %True once you enter the bibliography

\usepackage{microtype}
\usepackage{lineno}  % for line numbering during review
\usepackage{xspace} % To avoid problems with missing or double spaces after
\usepackage{caption} %these three command get the figure and table captions automatically small

\usepackage{graphicx}  % to include figures (can also use other packages)
\usepackage{color}
\usepackage{colortbl}
\graphicspath{{./figs/}} % Make Latex search fig subdir for figures

\usepackage{amsmath} % Adds a large collection of math symbols
\usepackage{amssymb}
\usepackage{amsfonts}
\usepackage{upgreek} % Adds in support for greek letters in roman typeset

\newcommand*\patchAmsMathEnvironmentForLineno[1]{%
\expandafter\let\csname old#1\expandafter\endcsname\csname #1\endcsname
\expandafter\let\csname oldend#1\expandafter\endcsname\csname
end#1\endcsname
 \renewenvironment{#1}%
   {\linenomath\csname old#1\endcsname}%
   {\csname oldend#1\endcsname\endlinenomath}%
}
\newcommand*\patchBothAmsMathEnvironmentsForLineno[1]{%
  \patchAmsMathEnvironmentForLineno{#1}%
  \patchAmsMathEnvironmentForLineno{#1*}%
}
\AtBeginDocument{%
\patchBothAmsMathEnvironmentsForLineno{equation}%
\patchBothAmsMathEnvironmentsForLineno{align}%
\patchBothAmsMathEnvironmentsForLineno{flalign}%
\patchBothAmsMathEnvironmentsForLineno{alignat}%
\patchBothAmsMathEnvironmentsForLineno{gather}%
\patchBothAmsMathEnvironmentsForLineno{multline}%
\patchBothAmsMathEnvironmentsForLineno{eqnarray}%
}

\usepackage{hyperref}    % Hyperlinks in references
\usepackage[all]{hypcap} % Internal hyperlinks to floats.

\usepackage{xspace} 
\usepackage{upgreek}

\def\lhcb {\mbox{LHCb}\xspace}

\def\belle  {\mbox{Belle}\xspace}

\def\cdf    {\mbox{CDF}\xspace}
\def\dzero  {\mbox{D0}\xspace}

\def\MagUp {\mbox{\em Mag\kern -0.05em Up}\xspace}

\ifthenelse{\boolean{uprightparticles}}%
{
 
 \def\Pgamma      {\ensuremath{\upgamma}\xspace}

 \def\Ppi         {\ensuremath{\uppi}\xspace}

 \def\PDelta      {\ensuremath{\Delta}\xspace}                 
 \def\PXi      {\ensuremath{\Xi}\xspace}                 
 \def\PLambda      {\ensuremath{\Lambda}\xspace}                 
 \def\PSigma      {\ensuremath{\Sigma}\xspace}                 
 \def\POmega      {\ensuremath{\Omega}\xspace}                 
 \def\PUpsilon      {\ensuremath{\Upsilon}\xspace}                 
 
 \def\PB      {\ensuremath{\mathrm{B}}\xspace}                 
                  
 \def\PD      {\ensuremath{\mathrm{D}}\xspace}

 \def\PK      {\ensuremath{\mathrm{K}}\xspace}

 \def\Pb      {\ensuremath{\mathrm{b}}\xspace}                 
 \def\Pc      {\ensuremath{\mathrm{c}}\xspace}

 \def\Pi      {\ensuremath{\mathrm{i}}\xspace}

 \def\Pp      {\ensuremath{\mathrm{p}}\xspace}

 \def\Ps      {\ensuremath{\mathrm{s}}\xspace}

}
{
 
 \def\Pgamma      {\ensuremath{\gamma}\xspace}

 \def\Ppi         {\ensuremath{\pi}\xspace}

 \mathchardef\PDelta="7101
 \mathchardef\PXi="7104
 \mathchardef\PLambda="7103
 \mathchardef\PSigma="7106
 \mathchardef\POmega="710A
 \mathchardef\PUpsilon="7107
                  
 \def\PB      {\ensuremath{B}\xspace}                 
                  
 \def\PD      {\ensuremath{D}\xspace}

 \def\PK      {\ensuremath{K}\xspace}

 \def\Pb      {\ensuremath{b}\xspace}                 
 \def\Pc      {\ensuremath{c}\xspace}

 \def\Pi      {\ensuremath{i}\xspace}

 \def\Pp      {\ensuremath{p}\xspace}

 \def\Ps      {\ensuremath{s}\xspace}

}

\makeatletter
\ifcase \@ptsize \relax% 10pt
  \newcommand{\miniscule}{\@setfontsize\miniscule{4}{5}}% \tiny: 5/6
\or% 11pt
  \newcommand{\miniscule}{\@setfontsize\miniscule{5}{6}}% \tiny: 6/7
\or% 12pt
  \newcommand{\miniscule}{\@setfontsize\miniscule{5}{6}}% \tiny: 6/7
\fi
\makeatother

\DeclareRobustCommand{\optbar}[1]{\shortstack{{\miniscule (\rule[.5ex]{1.25em}{.18mm})}
  \\ [-.7ex] $#1$}}

\def\g      {{\ensuremath{\Pgamma}}\xspace}

\def\squark    {{\ensuremath{\Ps}}\xspace}

\def\cquark    {{\ensuremath{\Pc}}\xspace}
\def\cquarkbar {{\ensuremath{\overline \cquark}}\xspace}

\def\bquark    {{\ensuremath{\Pb}}\xspace}

\def\pion   {{\ensuremath{\Ppi}}\xspace}
\def\piz    {{\ensuremath{\pion^0}}\xspace}

\def\pip    {{\ensuremath{\pion^+}}\xspace}
\def\pim    {{\ensuremath{\pion^-}}\xspace}

\def\kaon    {{\ensuremath{\PK}}\xspace}
  \def\Kbar    {{\kern 0.2em\overline{\kern -0.2em \PK}{}}\xspace}

\def\KorKbar    {\kern 0.18em\optbar{\kern -0.18em K}{}\xspace}

\def\Kp      {{\ensuremath{\kaon^+}}\xspace}
\def\Km      {{\ensuremath{\kaon^-}}\xspace}

  \def\Dbar    {{\kern 0.2em\overline{\kern -0.2em \PD}{}}\xspace}
\def\D       {{\ensuremath{\PD}}\xspace}

\def\DorDbar    {\kern 0.18em\optbar{\kern -0.18em D}{}\xspace}
\def\Dz      {{\ensuremath{\D^0}}\xspace}

\def\Dp      {{\ensuremath{\D^+}}\xspace}
\def\Dm      {{\ensuremath{\D^-}}\xspace}
\def\Dpm     {{\ensuremath{\D^\pm}}\xspace}

\def\Dstarp  {{\ensuremath{\D^{*+}}}\xspace}
\def\Dstarm  {{\ensuremath{\D^{*-}}}\xspace}

\def\Dstarmp {{\ensuremath{\D^{*\mp}}}\xspace}

\def\Dsp     {{\ensuremath{\D^+_\squark}}\xspace}
\def\Dsm     {{\ensuremath{\D^-_\squark}}\xspace}
\def\Dspm    {{\ensuremath{\D^{\pm}_\squark}}\xspace}
\def\Dsmp    {{\ensuremath{\D^{\mp}_\squark}}\xspace}

\def\Dssp    {{\ensuremath{\D^{*+}_\squark}}\xspace}
\def\Dssm    {{\ensuremath{\D^{*-}_\squark}}\xspace}
\def\Dsspm   {{\ensuremath{\D^{*\pm}_\squark}}\xspace}
\def\Dssmp   {{\ensuremath{\D^{*\mp}_\squark}}\xspace}

\def\B       {{\ensuremath{\PB}}\xspace}
\def\Bbar    {{\ensuremath{\kern 0.18em\overline{\kern -0.18em \PB}{}}}\xspace}

\def\BorBbar    {\kern 0.18em\optbar{\kern -0.18em B}{}\xspace}

\def\Bpm     {{\ensuremath{\B^\pm}}\xspace}

\def\Bd      {{\ensuremath{\B^0}}\xspace}
\def\Bs      {{\ensuremath{\B^0_\squark}}\xspace}
\def\Bsb     {{\ensuremath{\Bbar{}^0_\squark}}\xspace}
\def\Bdb     {{\ensuremath{\Bbar{}^0}}\xspace}

  \def\Y#1S{\ensuremath{\PUpsilon{(#1S)}}\xspace}% no space before {...}!

\def\proton      {{\ensuremath{\Pp}}\xspace}

\def\Lz          {{\ensuremath{\PLambda}}\xspace}
\def\Lbar        {{\ensuremath{\kern 0.1em\overline{\kern -0.1em\PLambda}}}\xspace}
\def\LorLbar    {\kern 0.18em\optbar{\kern -0.18em \PLambda}{}\xspace}

\def\Lb      {{\ensuremath{\Lz^0_\bquark}}\xspace}

\def\Lc      {{\ensuremath{\Lz^+_\cquark}}\xspace}

\def\BF         {{\ensuremath{\mathcal{B}}}\xspace}

\newcommand{\decay}[2]{\ensuremath{#1\!\to #2}\xspace}         % {\Pa}{\Pb \Pc}

\def\to                 {\ensuremath{\rightarrow}\xspace}

\def\CP                {{\ensuremath{C\!P}}\xspace}

\newcommand{\DGs}{{\ensuremath{\Delta\Gamma_{\squark}}}\xspace}

\newcommand{\Gs}{{\ensuremath{\Gamma_{\squark}}}\xspace}

\newcommand{\DGsGs}{{\ensuremath{\Delta\Gamma_{\squark}/\Gamma_{\squark}}}\xspace}

\def\AT#1     {\ensuremath{A_{\mathrm{T}}^{#1}}\xspace}           % 2

\def\C#1      {\ensuremath{\mathcal{C}_{#1}}\xspace}                       % 9
\def\Cp#1     {\ensuremath{\mathcal{C}_{#1}^{'}}\xspace}                    % 7
\def\Ceff#1   {\ensuremath{\mathcal{C}_{#1}^{\mathrm{(eff)}}}\xspace}        % 9  
\def\Cpeff#1  {\ensuremath{\mathcal{C}_{#1}^{'\mathrm{(eff)}}}\xspace}       % 7
\def\Ope#1    {\ensuremath{\mathcal{O}_{#1}}\xspace}                       % 2
\def\Opep#1   {\ensuremath{\mathcal{O}_{#1}^{'}}\xspace}                    % 7

\newcommand{\tev}{\ifthenelse{\boolean{inbibliography}}{\ensuremath{~T\kern -0.05em eV}\xspace}{\ensuremath{\mathrm{\,Te\kern -0.1em V}}}\xspace}
\newcommand{\gev}{\ensuremath{\mathrm{\,Ge\kern -0.1em V}}\xspace}
\newcommand{\mev}{\ensuremath{\mathrm{\,Me\kern -0.1em V}}\xspace}
\newcommand{\kev}{\ensuremath{\mathrm{\,ke\kern -0.1em V}}\xspace}
\newcommand{\ev}{\ensuremath{\mathrm{\,e\kern -0.1em V}}\xspace}
\newcommand{\gevc}{\ensuremath{{\mathrm{\,Ge\kern -0.1em V\!/}c}}\xspace}
\newcommand{\mevc}{\ensuremath{{\mathrm{\,Me\kern -0.1em V\!/}c}}\xspace}
\newcommand{\gevcc}{\ensuremath{{\mathrm{\,Ge\kern -0.1em V\!/}c^2}}\xspace}
\newcommand{\gevgevcccc}{\ensuremath{{\mathrm{\,Ge\kern -0.1em V^2\!/}c^4}}\xspace}
\newcommand{\mevcc}{\ensuremath{{\mathrm{\,Me\kern -0.1em V\!/}c^2}}\xspace}

\def\mum  {\ensuremath{{\,\upmu\mathrm{m}}}\xspace}

\def\invfb   {\ensuremath{\mbox{\,fb}^{-1}}\xspace}

\newcommand{\stat}{\ensuremath{\mathrm{\,(stat)}}\xspace}
\newcommand{\syst}{\ensuremath{\mathrm{\,(syst)}}\xspace}

\def\gsim{{~\raise.15em\hbox{$>$}\kern-.85em
          \lower.35em\hbox{$\sim$}~}\xspace}
\def\lsim{{~\raise.15em\hbox{$<$}\kern-.85em
          \lower.35em\hbox{$\sim$}~}\xspace}

\def\ptot       {\mbox{$p$}\xspace}
\def\pt         {\mbox{$p_{\mathrm{ T}}$}\xspace}

\def\evtgen     {\mbox{\textsc{EvtGen}}\xspace}

\def\geant      {\mbox{\textsc{Geant4}}\xspace}

\def\photos     {\mbox{\textsc{Photos}}\xspace}

\def\pythia     {\mbox{\textsc{Pythia}}\xspace}

\def\tell1  {TELL1\xspace}
\def\ukl1   {UKL1\xspace}

\usepackage{cite} % Allows for ranges in citations
\usepackage{mciteplus}

\begin{document}

%%%%%%%%%%%%%%%%%%%%%%%%%
%%%%% Title     %%%%%%%%%
%%%%%%%%%%%%%%%%%%%%%%%%%
\renewcommand{\thefootnote}{\fnsymbol{footnote}}
\setcounter{footnote}{1}

% $Id: title-LHCb-PAPER.tex 78711 2015-08-06 07:54:32Z apuignav $
% ===============================================================================
% Purpose: LHCb-PAPER journal paper title page template
% Author: 
% Created on: 2010-09-25
% ===============================================================================

%%%%%%%%%%%%%%%%%%%%%%%%%
%%%%%  TITLE PAGE  %%%%%%
%%%%%%%%%%%%%%%%%%%%%%%%%
\begin{titlepage}
\pagenumbering{roman}

% Header ---------------------------------------------------
\vspace*{-1.5cm}
\centerline{\large EUROPEAN ORGANIZATION FOR NUCLEAR RESEARCH (CERN)}
\vspace*{1.5cm}
\noindent
\begin{tabular*}{\linewidth}{lc@{\extracolsep{\fill}}r@{\extracolsep{0pt}}}
%\ifthenelse{\boolean{pdflatex}}% Logo format choice
%{\vspace*{-2.7cm}\mbox{\!\!\!\includegraphics[width=.14\textwidth]{lhcb-logo.pdf}} & &}%
%{\vspace*{-1.2cm}\mbox{\!\!\!\includegraphics[width=.12\textwidth]{lhcb-logo.eps}} & &}%
\\
 & & CERN-EP-2016-028 \\  % ID 
 & & LHCb-PAPER-2015-053 \\  % ID 
 & & April 25, 2016 \\ % Date - Can also hardwire e.g.: 23 March 2010
 & & \\
% not in paper \hline
\end{tabular*}

\vspace*{2.0cm}

% Title --------------------------------------------------
{\normalfont\bfseries\boldmath\LARGE
\begin{center}
  Measurement of the \BsDsDs branching fractions
\end{center}
}

\vspace*{0.5cm}

% Authors -------------------------------------------------
\begin{center}
%In the footnote, replace 'paper' by 'letter' in case of submission to PRL or PLB 
The LHCb collaboration\footnote{Authors are listed at the end of this paper.}
\end{center}

\vspace{\fill}
%\vspace{5mm}

% Abstract -----------------------------------------------
\begin{abstract}
\noindent  The branching fraction of the decay \BsDsDs is
  measured using $pp$ collision data corresponding to an
  integrated luminosity of $1.0$\invfb, collected using the \lhcb
  detector at a centre-of-mass energy of $7$\tev. It is found to be
\begin{align*}
    \BF(B_{s}^{0}\rightarrow~D_{s}^{(\ast)+}D_{s}^{(\ast)-}) = (3.05 \pm 0.10 \pm 0.20 \pm 0.34)\%,
\end{align*}
  \noindent
  where the uncertainties are statistical, systematic, and due to the
  normalisation channel, respectively.
  The branching fractions of the individual decays corresponding to the presence of one or
  two \Dsspm are also measured. 
  The individual branching fractions are found to be
\begin{align*}
    \BF(\Bs\to\Dsspm\Dsmp) = (1.35 \pm 0.06 \pm 0.09 \pm 0.15)\%,\\
    \BF(\Bs\to\Dssp\Dssm) = (1.27 \pm 0.08 \pm 0.10 \pm 0.14)\%.
\end{align*}

%\vspace{6mm}

\noindent
All three results are the most precise determinations to date.

\end{abstract}

\vspace*{0.5cm}

\begin{center}
  Published in Phys.~Rev.~D 92, 092008 on 20th May 2016
\end{center}

\vspace{\fill}

{\footnotesize 
\centerline{\copyright~CERN on behalf of the \lhcb collaboration, licence \href{http://creativecommons.org/licenses/by/4.0/}{CC-BY-4.0}.}}
\vspace*{2mm}

\end{titlepage}

%%%%%%%%%%%%%%%%%%%%%%%%%%%%%%%%
%%%%%  END OF TITLE PAGE  %%%%%%
%%%%%%%%%%%%%%%%%%%%%%%%%%%%%%%%

%  empty page follows the title page ----
\newpage
\setcounter{page}{2}
\mbox{~}

\cleardoublepage

\renewcommand{\thefootnote}{\arabic{footnote}}
\setcounter{footnote}{0}

%%%%%%%%%%%%%%%%%%%%%%%%%
%%%%% Main text %%%%%%%%%
%%%%%%%%%%%%%%%%%%%%%%%%%

\pagestyle{plain} % restore page numbers for the main text
\setcounter{page}{1}
\pagenumbering{arabic}

\section{Introduction}
\label{sec:Intro}

Because of $\Bs-\Bsb$ oscillations, the mass and flavour eigenstates of the \Bs system do not coincide.
The \Bs meson mass eigenstates have a relative decay width difference \DGsGs, where
\DGs (\Gs) is the difference (average) of the decay widths between the heavy and light states. 
The relative decay width difference is one of the key parameters of the \Bs system, and its precise determination
allows stringent tests of the flavour sector of the Standard Model.

Under certain theoretical assumptions, \BsDsDs decays were thought to saturate the \CP-even contribution
to \DGs, and therefore the branching fraction of \BsDsDs was used as a means of approximating \DGsGs~\cite{Aleksan1993567}.
This approximation is now considered to be a poor one~\cite{Lenz:2012mb}, as the decay modes containing at least one \Dsspm have a 
non-negligible \CP-odd component, and other three-body \Bs decays can contribute to the value of \DGs at a similar level as \BsDsDs decays.
A detailed discussion of theoretical
predictions of the \BsDsDs branching fractions, and the predicted contribution of other modes to the
value of \DGsGs, is given in Ref.~\cite{Chua:2011er}.

In a more general context, since the branching fraction of
\BsDsDs decays is one of the dominant contributions
to the total inclusive \decay{\bquark}{\cquark\cquarkbar\squark} branching fraction, its
precise measurement is an important ingredient in model-independent
searches for physics beyond the Standard Model in $B$ meson decays~\cite{Krinner201331}. The most recent
measurements are provided by the \belle~\cite{Esen:2012yz}, \cdf~\cite{Aaltonen:2012mg}, and \dzero~\cite{Abazov:2008ig} collaborations
who obtain, respectively,
\begin{align*}
\BF(\BsDsDs)& = (4.32_{-0.39 -1.03}^{+0.42 +1.04})\%,\\
\BF(\BsDsDs)& = (3.38 \pm 0.25 \pm 0.30 \pm 0.56)\%,\\
\BF(\BsDsDs)& = (3.5 \pm 1.0 \pm 1.1)\%.
\end{align*}

The data used in the analysis presented in this paper correspond to an integrated luminosity of $1.0\invfb$, 
collected by the \lhcb experiment during the 2011 run period.
The branching fraction of the full $\BsDsDs$ decay is determined
relative to the $\Bd\to\Dsp\Dm$ decay, which has a similar final state and a precisely measured branching
fraction. The charm daughters are reconstructed using the $\Dsp\to\Kp\Km\pip$ and $\Dm\to\Kp\pim\pim$ final states.
Throughout the paper, unless stated otherwise, charge-conjugate modes are implied and summed over.
The branching fraction ratio is determined as
\begin{equation}
\label{eq:BR}
\frac{\BF(\BsDsDs)}{\BF(B^{0}\rightarrow \Dsp\Dm)} = \frac{f_{d}}{f_{s}} \cdot 
\frac{\epsilon^{B^{0}}}{\epsilon^{B_{s}^{0}}} \cdot \frac{\BF(D^{-}\rightarrow~K^{+}\pi^{-}\pi^{-})}{\BF(D_{s}^{+}\rightarrow~K^{-}K^{+}\pi^{+})}
\cdot \frac{N_{B_{s}^{0}}}{N_{B^{0}}},
\end{equation}
\noindent where $f_{d}$ ($f_{s}$) is the fraction of \Bdb (\Bsb) mesons produced in the fragmentation of a \bquark quark,
$\epsilon^{B^{0}}/\epsilon^{B_{s}^{0}}$ is the relative efficiency of the \Bd to the \Bs
selections, $\BF(D^{-}\rightarrow~K^{+}\pi^{-}\pi^{-})$ and $\BF(D_{s}^{+}\rightarrow~K^{-}K^{+}\pi^{+})$ are the branching fractions
of the charm daughter decays, and $N_{B_{s}^{0}}/N_{B^{0}}$ is the relative yield of \Bs and \Bd candidates.

The branching fraction of the exclusive $\Bs\to\Dssp\Dssm$ decay is determined in the same way, along with the 
branching fraction of $\Bs\to\Dsspm\Dsmp$. 
The branching fraction of the $\Bs\to\Dsp\Dsm$ decay
has been previously measured by \lhcb using the same data as this analysis~\cite{LHCb-PAPER-2012-050}, and is therefore not determined in this study. 
However, the selection efficiency and yield in the $\Bs\to\Dsp\Dsm$ channel are determined in this analysis,
as both are needed for the calculation of the total \BsDsDs selection efficiency.

\section{Detector and simulation}
\label{sec:Detector}

The \lhcb detector~\cite{Alves:2008zz,LHCb-DP-2014-002} is a single-arm forward
spectrometer covering the \mbox{pseudorapidity} range $2<\eta <5$,
designed for the study of particles containing \bquark or \cquark
quarks. The detector includes a high-precision tracking system
consisting of a silicon-strip vertex detector surrounding the $pp$
interaction region, a large-area silicon-strip detector located
upstream of a dipole magnet with a bending power of about
$4{\mathrm{\,Tm}}$, and three stations of silicon-strip detectors and straw
drift tubes placed downstream of the magnet.
The tracking system provides a measurement of momentum, \ptot, of charged particles with
a relative uncertainty that varies from 0.5\% at low momentum to 1.0\% at 200\gevc.
The minimum distance of a track to a primary vertex, the impact parameter, is measured with a resolution of $(15+29/\pt)\mum$,
where \pt is the component of the momentum transverse to the beam, in\,\gevc.
Different types of charged hadrons are distinguished using information
from two ring-imaging Cherenkov detectors. 
Photons, electrons and hadrons are identified by a calorimeter system consisting of
scintillating-pad and preshower detectors, an electromagnetic
calorimeter and a hadronic calorimeter. Muons are identified by a
system composed of alternating layers of iron and multiwire
proportional chambers.
The event selection is performed in two stages, with an initial online selection followed by a
tighter offline selection.
The online event selection is performed by a trigger~\cite{LHCb-DP-2012-004}, which consists of a
hardware stage, based on information from the calorimeter and muon
systems, followed by a software stage, which performs a full event reconstruction.

In the simulation, $pp$ collisions are generated using
\pythia6~\cite{Sjostrand:2006za} with a specific \lhcb
configuration~\cite{LHCb-PROC-2010-056}.  Decays of hadronic particles
are described by \evtgen~\cite{Lange:2001uf}, in which final-state
radiation is generated using \photos~\cite{Golonka:2005pn}. The
interaction of the generated particles with the detector, and its response,
are implemented using the \geant toolkit~\cite{Allison:2006ve, *Agostinelli:2002hh} as described in
Ref.~\cite{LHCb-PROC-2011-006}.

\section{Signal selection}
\label{sec:selection}
The \Dssp meson decays to a \Dsp meson and either a photon or a neutral pion $(93.5\pm0.7)\%$ and $(5.8\pm0.7)\%$ of the time, 
respectively, nearly saturating the total branching fraction. The remainder of the decays are ignored in this analysis. 
%There are also a small number of decays to $\Dsp\ep\en$, but these are disregarded in this analysis.
Neither of the neutral particles is reconstructed in the
decay chain, and the individual $\Bs\to\Dsspm\Dsmp$ and $\Bs\to\Dssp\Dssm$ decays are identified through the reconstructed invariant
mass of the $\Dsp\Dsm$ system. 
The individual peaks from $\Bs\to\Dsspm(\to\Dspm\g)\Dsmp$ and $\Bs\to\Dsspm(\to\Dspm\piz)\Dsmp$ are not resolved.
Therefore the reconstructed $\Dsp\Dsm$ mass distribution has three separate peaks, corresponding to decays containing zero, one, or two \Dsspm particles.

At the hardware trigger stage,
events are required to have a muon with high \pt or a hadron, photon or electron with high transverse energy in the calorimeters. 
For hadrons, the transverse energy threshold is $3.5\gev$. 
Candidate \Bs and \Bd mesons are used in the analysis if at least one of the associated tracks is selected by
the hardware trigger, 
or if the event is triggered independently of the particles in the signal decay.
The software
trigger considers all charged particles with $\pt>500\mevc$ and
constructs two-, three-, or four-track secondary vertices which require a significant displacement from the primary 
$pp$ interaction vertices. At least one charged particle
must have a transverse momentum $\pt > 1.7\gevc$ and be
inconsistent with originating from a primary vertex.
A multivariate algorithm~\cite{BBDT} is used for
the identification of secondary vertices consistent with the decay
of a \bquark hadron. The selection to this point is hereafter referred to as the initial selection.

Signal \Bs and normalisation $\Bd\to\Dsp\Dm$ candidates are required to satisfy a number of additional conditions in order to be included in the final
samples. 
Kaons and pions are required to be identified by the particle identification (PID) system. 
All \Dsp and \Dm candidates must have an invariant mass within $\pm30\mevcc$ of their known values~\cite{PDG2014}. 
Signal \Bs candidates are required to have a reconstructed mass
in the range $4750-5800\mevcc$, whereas \Bd candidates must have a mass in the range $5050-5500\mevcc$.
After these requirements are applied there
are still contributions from other \bquark-hadron decays into final states with two charm particles. The decays $\Lb\to\Lc(\to\proton\Km\pip)\Dsm$, where
the \proton is misidentified as a \Kp, and $\Bd\to\Dsp\Dm(\to\Kp\pim\pim)$, where a \pim is misidentified as a
\Km, result in background contamination in the signal channel, while the decay $\Bs\to\Dsp\Dsm$ 
contributes to the background in the normalisation channel if the \Kp
in $\Dsp\to\Kp\Km\pip$ is misidentified as a \pip. As these backgrounds accumulate in reconstructed mass close to the signal peaks,
candidates consistent with any one of these background decay hypotheses
are rejected in the selection by applying a veto based on the invariant mass of the candidate under 
the alternative particle type hypotheses. Candidate \Dsp mesons
are vetoed if they have a reconstructed mass in the range $2271-2301\mevcc$ when the \Kp candidate is assumed 
to be a proton, or a mass in the range $1835-1905\mevcc$ 
when the \Kp candidate is assigned the \pip mass. Candidate \Dm mesons are vetoed if they have a reconstructed mass in the 
range $1950-1990\mevcc$ when a \pim candidate is assigned the kaon mass.
In a simulated sample of $\Bd\to\Dsp\Dm$ decays, $17.7\%$ of the events meet all of the \BsDsDs selection criteria before the \Dpm veto
is applied. After the veto, only $0.05\%$ of the simulated $\Bd\to\Dsp\Dm$ sample still pass the full \BsDsDs selection.
The decay $\Bd\to\Dsp\Dsm$ and three-body $\Bpm\to D_{(s)}^{+}D_{(s)}^{-}h^{\pm}$ decays, where $h$ is either a kaon or pion,
are examined as other potential background sources, but
are all disregarded because of either a small selection efficiency or small branching fraction relative to the signal channels.

In order to further improve the purity of the signal and normalisation samples, a boosted decision tree (BDT) classifier is used to
distinguish real $B^{0}_{(s)}$ decays from combinatorial background~\cite{Breiman}. The BDT is trained using the AdaBoost algorithm~\cite{AdaBoost} to distinguish 
simulated \Bs signal decays from background candidates obtained from mass sidebands in the data.
Background candidates must contain a \Bs candidate with a mass greater than $5600\mevcc$ and two \Dspm candidates
with masses less than $1930\mevcc$ or greater than $2010\mevcc$. The set of 14 variables used as input to the BDT exploits the topology of the 
\Bs decay chain and includes the transverse momentum of the \Bs candidate and of the two \Dspm daughters, as well as the product of the absolute transverse momenta
of the pions and kaons produced in the decay of each \Dspm. The decay times of the two \Dspm candidates with respect to the primary vertex 
and variables related to the consistency of the \Bs and of the two \Dspm to come from the primary vertex are also used.
The optimal BDT requirement is chosen to maximise the value of $N_{s}/\sqrt{N_{s}+N_{b}}$, 
where $N_{s}$ is the total number of signal candidates matching any of the three exclusive decays in \BsDsDs and
$N_{b}$ is the total number of combinatorial background events as taken from the fit. The same BDT classifier and selection criteria 
are also applied to the normalisation sample.

\begin{table}[t]
\small
\begin{center}
\caption{\label{tab:Efficiencies}
    \small %captions should be a little bit smaller than main text
    Efficiencies of the various selection criteria for the three individual channels of \BsDsDs, and for $\Bd\to \Dsp\Dm$. Each efficiency is presented
    relative to the previous cut and measured using simulated events, except for the PID efficiency which is obtained from data. 
    The \Dsp veto was only applied to the normalisation mode, $\Bd\to\Dsp\Dm$.}

\begin{tabular}{@{}c|r@{ $\pm$ }l@{  }|r@{ $\pm$ }l@{ }|r@{ $\pm$ }l@{ }|r@{\ $\pm$\ }l@{ }}
%\begin{tabular}{@{}c|r@{ $\pm$ }l@{  }r@{ $\pm$ }l@{ }r@{ $\pm$ }l@{ }r@{\ $\pm$\ }l@{ }}
    \hline
     & \multicolumn{8}{c}{Selection efficiency (\%)} \Tstrut \\
    \multicolumn{1}{c|}{Selection} & \multicolumn{2}{c}{$\Bs\to\Dsp\Dsm$} & \multicolumn{2}{c}{$\Bs\to \Dsspm\Dsmp$} & \multicolumn{2}{c}{$\Bs\to\Dssp\Dssm$}  & \multicolumn{2}{c}{$\Bd\to \Dsp\Dm$} \Tstrut   \\ 
    \hline
    Reconstruction & 0.1184 & 0.0003 & 0.1127 & 0.0005 & 0.1061 & 0.0005 & 0.1071 & 0.0002 \Tstrut \\
    Initial selection & 1.362 & 0.008 & 1.250 & 0.010 & 1.100 & 0.010 & 1.416 & 0.009 \\
    Mass requirements & 89.4 & 0.6 & 87.8 & 1.0 & 88.3 & 1.0 & 88.5 & 0.6 \\
    BDT & 97.9 & 0.7 & 96.6 & 1.1 & 96.7 & 1.1 & 97.6 & 0.7 \\
    \Dp veto & 48.7 & 0.5 & 50.3 & 0.8 & 48.9 & 0.8 & 68.7 & 0.6 \\
    \Dsp veto & \multicolumn{2}{c|}{$-$} & \multicolumn{2}{c|}{$-$} & \multicolumn{2}{c|}{$-$} & 64.8 & 0.7 \\
    \Lc veto & 96.3 & 1.0 & 96.3 & 1.6 & 95.9 & 1.6 & 98.2 & 0.8 \\
    Trig. requirement & 96.6 & 0.7 & 96.7 & 1.1 & 96.6 & 1.1 & 96.8 & 0.7 \\
    PID requirements & 82.4 & 0.2 & 82.4 & 0.2 & 82.4 & 0.2 & 84.2 & 0.1 \\
    \hline
    Total & 0.0527 & 0.0067 & 0.0460 & 0.0095 & 0.0372 & 0.0081 & 0.0467 & 0.0060 \Tstrut \\
    \hline
\end{tabular}

\end{center}
\end{table}

The efficiencies of the selection criteria in both the signal and normalisation channels are listed in Table~\ref{tab:Efficiencies}.
The efficiencies of the background vetoes,
trigger, reconstruction, and BDT selection are determined using simulated signal samples. The efficiencies of identifying \Kp and \pip mesons are determined
using a calibration data sample of $\Dstarp\to\Dz(\to\Km\pip)\pip$ decays, with kinematic quantities reweighted to match those of the signal candidates. 
The efficiency of the PID selection is found to be $(82.4\pm0.2)\%$ for signal \Bs decays and
$(84.2\pm0.1)\%$ for \Bd decays. The efficiency of the full \BsDsDs decay is determined by calculating a weighted average of the individual signal channels, with
weights given by the relative yields in data. The relative efficiencies of the \Bd decay to the three individual channels and the full decay are given 
in Table~\ref{tab:RelEffs}.

\begin{table}[t]
\begin{center}
\caption{\label{tab:RelEffs}
    \small %captions should be a little bit smaller than main text
    Efficiency of the normalisation channel $\Bd\to\Dsp\Dm$ relative to the signal decays.}
\begin{tabular}{l|c}
	      \hline
              \multicolumn{1}{c|}{Channel} & $\epsilon^{\Bd}/\epsilon^{\Bs}$ \Tstrut \\
              \hline
              $\Bs\to\Dsp\Dsm$ & $0.89\pm0.02$\Tstrut\\
              $\Bs\to\Dsspm\Dsmp$ & $1.02\pm0.03$ \\
              $\Bs\to\Dssp\Dssm$ & $1.26\pm0.03$ \\
              $\BsDsDs$& $1.06\pm0.02$ \\
              \hline
              \end{tabular}
\end{center}
\end{table}

\section{Signal and background shapes}
\label{sec:PDFs}

The \Bs and \Bd yields in the signal channels and the normalisation mode are extracted by performing a
three-dimensional extended unbinned
maximum likelihood fit to the mass distributions of the $B^{0}_{(s)}$ meson and the two charm daughters.

In order to determine the yields for the individual signal peaks, the \Bs candidate mass distribution
in each channel is modelled using simulated signal events. The $\Bs\to\Dsp\Dsm$ peak is parameterised as the sum of a Crystal Ball
function~\cite{Skwarnicki:1986xj} and a Gaussian function. The tail parameters of the Crystal Ball function, 
the ratio of the width of its Gaussian core to the width of the Gaussian function, 
and the relative weight of each function in the full distribution, are taken from simulation. 
The mean and width of the Gaussian core are allowed to float.
The two \Dspm distributions are also parameterised using this model, with all shape parameters fixed
to the values found in simulation.

Because of the kinematic differences between the $\Dssp\to\Dsp\g$ and $\Dssp\to\Dsp\piz$ decays, the peak 
of the $\Bs\to\Dsspm\Dsmp$ mass distribution is parameterised by a superposition of two Gaussian functions. The individual mean values, the ratio of the widths, 
and the fraction of each Gaussian function in the full distribution are fixed to values taken from simulation.
The peak corresponding to $\Bs\to\Dssp\Dssm$ decays is modelled using a single Gaussian function, with the mean
fixed to the value found from simulated events.

There is also a component in the fit to describe the presence of background decays of the form $\Bs\to\DsJp\Dsm$, where the $\DsJp$
can be either a \Dsstst or a \Dsststz meson that
decays to a \Dsp along with some combination of photons and neutral or charged pions. As some decay
products are missed, this background is present only at the low mass region of the signal distribution. The shape of the
distribution is determined by fitting to $\Bs\to\Dsstst\Dsm$ simulated events, as the contribution from \Dsstst is currently the best understood among the
$\DsJp$ decays. It is found to be well modelled by an Argus function~\cite{Albrecht1990278}, all shape parameters for which are
fixed to the values found in simulation.

The combinatorial background shape in the \Bs candidate mass distribution is parameterised by a second-order polynomial, 
and the model is validated with candidates passing a wrong-sign version of the
selection. The wrong-sign selection is identical to the signal selection but instead looks for events
containing two $\Dsp$ mesons.
The parameters of the combinatorial background distribution are allowed to float in the full fit to data, and are 
found to be compatible with the values obtained from the fit to the wrong-sign sample. The combinatorial background shape in the \Dspm distribution
is determined using events taken from the high-mass sideband region of the \Bs distribution, and is found to be consistent with a first-order
polynomial. 
The impact of adding a small Gaussian contribution to account for the presence of real \Dspm mesons in the combinatorial background
was found to be minimal, with the observed deviations from the nominal signal yields being smaller than the statistical uncertainty
in each case.

The \Bd distribution is modelled using the same parameterisation as for the full \Bs distribution, with one exception. 
The peak where either the \Dsp or \Dm comes from the decay of
an excited state is modelled by a superposition of three Gaussian functions, rather than the two-Gaussian model used in the \Bs case, 
to account for the difference in
distributions from \Dssp and \Dstarm decays, as the \Dstarm decay contains a \piz in the final state more frequently than \Dssp decays. There is also
a small contribution from the decay $\Bs\to\Dsm\Dp$, which is modelled with the same distribution as for the signal \Bs candidates.

\section{Fit results}
\label{sec:results}

\begin{figure}[t]
  \begin{center}
    \includegraphics[width=0.75\linewidth]{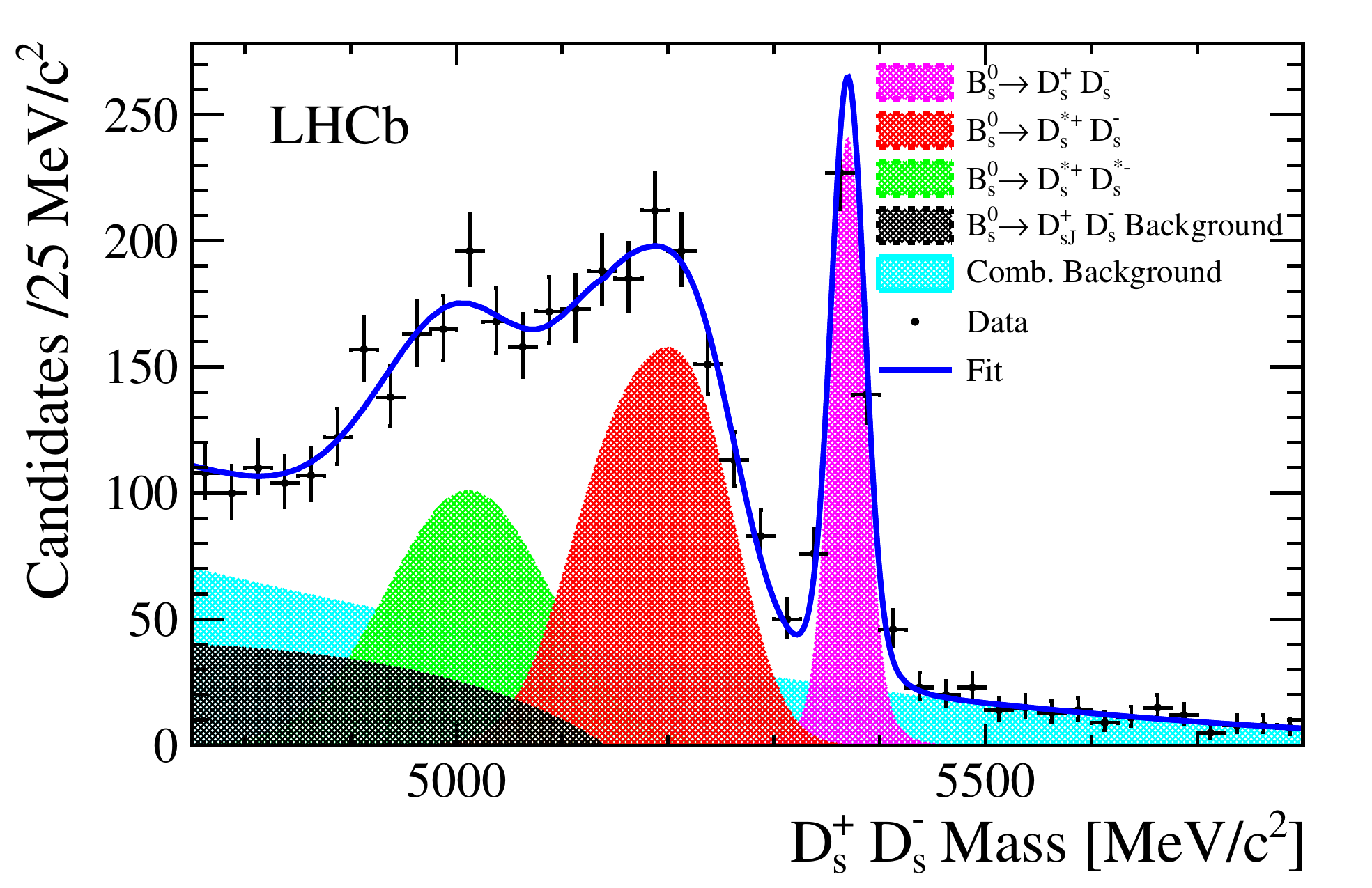}
    \caption{
    \small %captions should be a little bit smaller than main text
    \label{fig:BsDataFits}Invariant mass distribution of the \BsDsDs candidates.
    Also shown is the fit function and the individual components of the fit model.}
  \end{center}
\end{figure}

\begin{figure}[t]
  \begin{center}
    \includegraphics[width=0.75\linewidth]{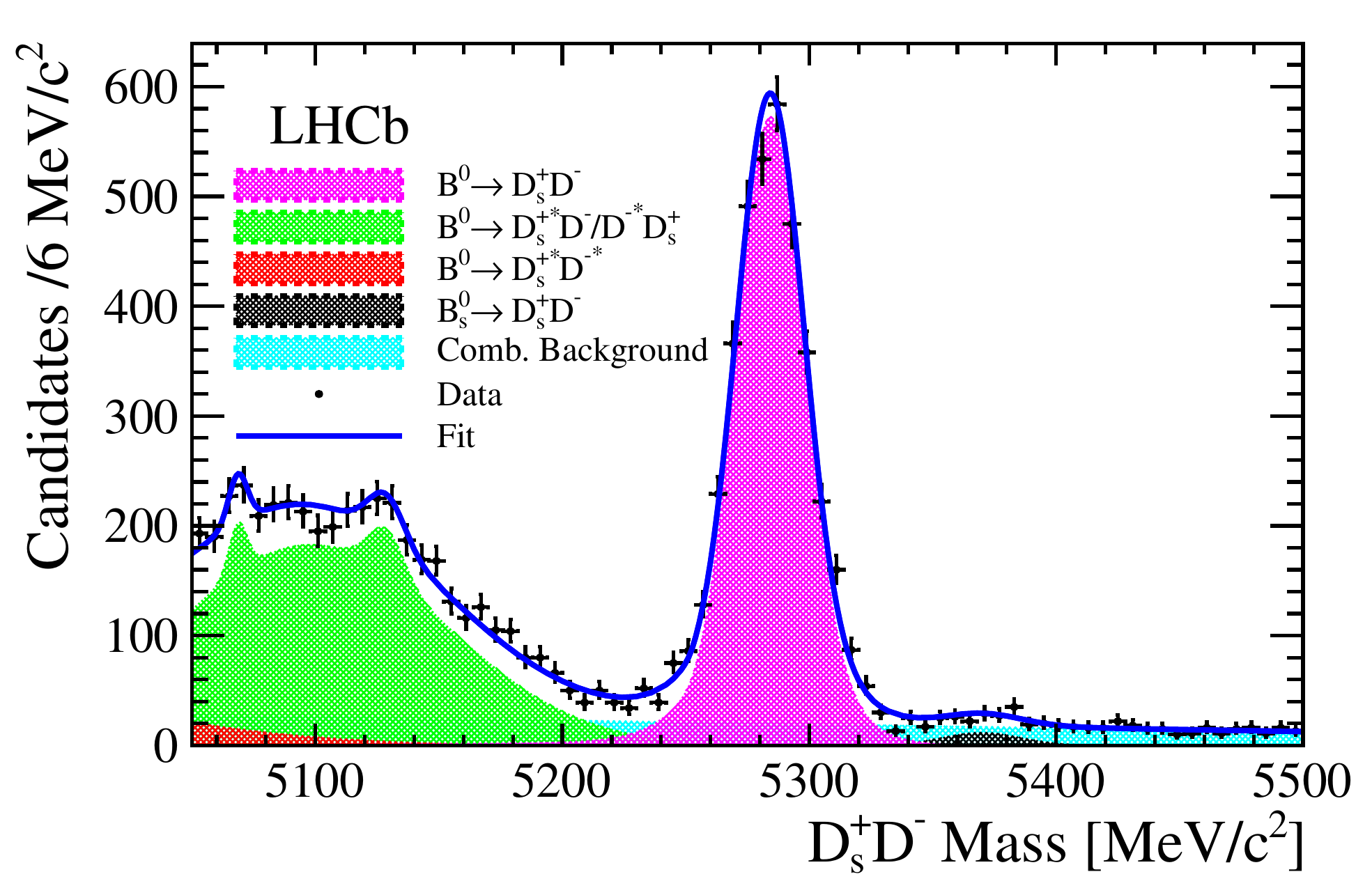}
    \caption{
    \small %captions should be a little bit smaller than main text
    \label{fig:BDataFits}Invariant mass distribution of the \BDsD candidates. 
    Also shown is the fit function and the individual components of the fit model.}
  \end{center}
\end{figure}

The fit to the signal data samples is shown in Fig.~\ref{fig:BsDataFits}, where the triple
peaked structure of the full decay is clearly visible.
The yields for the individual signal channels and the two backgrounds are given
in Table~\ref{tab:BsDsDsYields}.
The total \BsDsDs yield is the sum of the individual signal channel yields, with the uncertainty calculated using
the correlation coefficients between the individual yields, and is found to be $2230\pm63$.
The full fit to the data sample for the normalisation mode is shown in Fig.~\ref{fig:BDataFits}, and the yields are given in
Table~\ref{tab:BDsDYields}. 
Almost all $\Bd\rightarrow\Dsspm\Dstarmp$ decays are reconstructed with a mass lower than the 5050\mevcc\ mass cut 
imposed on the \Bd\ candidates. There is thus a relatively small yield from this channel.
Only the main $\Bd\rightarrow\Dsp\Dm$ peak is used for normalisation purposes.

\begin{table}[t]
\begin{center}
\caption{\label{tab:BsDsDsYields}
    \small %captions should be a little bit smaller than main text
    The yields extracted from the fit to the \BsDsDs candidate sample.}
\begin{tabular}{@{}l|r@{ $\pm$ }l@{ }}
    \hline
    \multicolumn{1}{c|}{Decay Mode}             & \multicolumn{2}{c}{Yield}   \Tstrut   \\ 
    \hline
    $\Bs\rightarrow\Dsp\Dsm$ & 412 & 23 \Tstrut\\
    $\Bs\rightarrow\Dsspm\Dsmp$ & 1032 & 39\\
    $\Bs\rightarrow\Dsspm\Dssmp$ & 786 & 48\\
    Combinatorial background & 1342 & 47\\
    $\Bs\rightarrow D_{s}(2460)^{\pm}\Dsmp$ & 432 & 42\\
    \hline
  \end{tabular}
\end{center}
\end{table}

\begin{table}[t]
\begin{center}
\caption{
    \small %captions should be a little bit smaller than main text
    The yields extracted from the fit to the \BDsD candidate sample.}
    \label{tab:BDsDYields}
\begin{tabular}{@{}l|r@{ $\pm$ }l@{}}
    \hline
    \multicolumn{1}{c|}{Decay Mode}             & \multicolumn{2}{c}{Yield}  \Tstrut    \\ 
    \hline
    $\Bd\rightarrow \Dsp\Dm$ & 3636 & 64 \Tstrut \\
    $\Bd\rightarrow \Dssp\Dm/\Dstarm\Dsp$ & 3579 & 110\\
    $\Bd\rightarrow \Dssp\Dstarm$ & 166 & 86\\
    $\Bs\rightarrow D_{s}^{\pm}D^{\mp}$ & 85 & 13\\
    Combinatorial background & 1542 & 56\\
    \hline
  \end{tabular}
\end{center}
\end{table}

\section{Systematic uncertainties}
\label{sec:systematics}

A number of systematic uncertainties affect the measurements of the ratios of branching fractions; the
sources and magnitudes of these uncertainties are summarised in Table~\ref{tab:totalsystematic}. The dominant
source of uncertainty for two of the three branching fractions comes from the $\bquark$ fragmentation fraction 
ratio, $f_{s}/f_{d} = 0.259\pm0.015$~\cite{LHCb-CONF-2013-011}. 
Part of the uncertainty on this ratio is due to the ratio of the charm branching fractions 
$\B(\Dsp\to\Kp\Km\pip)/\B(\Dm\to\Kp\pim\pim) = 0.594\pm0.020$~\cite{LHCb-CONF-2013-011}, the inverse of which is used in the measurements presented
in this paper, as shown in Eq.~\ref{eq:BR}. 
With the two values from Ref.~\cite{LHCb-CONF-2013-011},
the part of the uncertainty on $f_{s}/f_{d}$ due to the charm branching fractions cancels, 
leading to a total uncertainty for the product $f_{d}/f_{s} \times \B(\Dm\to\Kp\pim\pim)/\B(\Dsp\to\Kp\Km\pip)$ of $4.7\%$.

The fit model used for the yield extraction is validated using pseudoexperiments and is found to be unbiased. The uncertainty due to
the imperfect knowledge of the shape of the full mass distribution is investigated by measuring
the yields using alternative models for each of the peaks. 
The $\Dsp\Dsm$ peak is modelled with an Apollonios function~\cite{Santos:2013gra} or a Cruijff function~\cite{delAmoSanchez:2010ae}, the $\Dsspm\Dsmp$ peak is modelled using a single
Gaussian function, and the $\Dssp\Dssm$ peak is modelled using a combination of two Gaussian functions. 
The \BDsD fit model uncertainty is assessed by modeling the $\Bd\to\Dsm\Dp$ peak with both an Apollonios function and a Cruijff function.
In all cases, the systematic uncertainty
is taken to be the RMS deviation of the sets of yields with respect to the nominal yields found using the standard fits. The $\Bd\to\Dsm\Dp$
uncertainty is added in quadrature to the signal channel uncertainties,
leading to a systematic uncertainty of $3.4\%$ for the $\Dsspm\Dsmp$ branching fraction ratio, $2.2\%$ for the $\Dssp\Dssm$ branching fraction ratio, 
and $2.2\%$ for the total \BsDsDs branching fraction ratio.

The uncertainty on the combinatorial background yield is determined by considering the differences when
instead fitting this background with an exponential function, and is of the order of $1.5\%$ for all of the branching fraction ratios.

The dominant uncertainty for the $\Bs\rightarrow\Dssp\Dssm$ decay channel results from the lack of knowledge of the $\Bs\to\DsJp\Dsm$
background decays. The shape of this background overlaps mostly with the $\Bs\rightarrow\Dssp\Dssm$ signal decays, and therefore the systematic
uncertainty due to this background shape is much larger for this channel ($5.0\%$) than for the other two exclusive branching fractions ($0.2\%-0.4\%$).
The uncertainty is measured by repeating the fit with the cut-off point of the Argus function 
varied from $5050\mevcc$ to $5200\mevcc$, where the upper limit is chosen in order to account for the presence of decays containing \Dsststz mesons.
The changes to the yields from the values found in the nominal fit are calculated in each case. 
The systematic uncertainty in each channel is then assigned as
the RMS of the full set of deviations. The
uncertainty on the overall branching fraction ratio is also determined in this way, and is found to be $1.9\%$.

The uncertainties on the overall efficiencies due to the limited 
size of the simulated samples are calculated individually for each channel. For the total
measurement, $\BF(\BsDsDs)$, a weighted average of the individual uncertainties is used, with weights proportional to the final yield values obtained from data. 
These uncertainties on the efficiencies are then propagated to the branching fraction ratios.

There is a systematic uncertainty arising from the calculation of the efficiencies of the PID cuts. The calibration of the data samples is 
performed in bins of momentum and pseudorapidity,
which results in an uncertainty on the calculated efficiencies 
owing to the finite size of the $\Dstarp\to\Dz\pip$ calibration samples
and the binning scheme used.
The uncertainty resulting from the calibration sample size and binning scheme is determined by
redoing the calibration using different binning schemes. 
Another systematic uncertainty is due to the presence
of a small combinatorial background component in the samples that are used to
determine the PID efficiencies.
The systematic uncertainty due to this
contamination is estimated by comparing the efficiencies found in data to those found when calibrating simulated signal
events.
The total uncertainties due to the PID efficiency calculation for the three branching fraction ratios presented in this paper
are shown in Table~\ref{tab:totalsystematic}. The value for \BsDsDs is again the weighted average of the contributing channels, with the uncertainty
for the $\Dsp\Dsm$ contribution being $1.1\%$.

The uncertainty of $1.5\%$ from the trigger response is assessed by considering variations in the response between data and
simulation. The individual uncertainties are combined in quadrature to give the total relative systematic uncertainties for each measurement given in
Table~\ref{tab:totalsystematic}.

\begin{table}[]
\begin{center}
\caption{\label{tab:totalsystematic}
    \small %captions should be a little bit smaller than main text
    Systematic uncertainties, in \% of the relevant branching fraction ratio, for the \BsDsDs branching fraction ratios.}  
\begin{tabular}{c|c|c|c}
    \hline
    Source  & $\Bs\rightarrow\Dsspm\Dsmp$ & $\Bs\rightarrow\Dssp\Dssm$ & \BsDsDs \Tstrut \\
    \hline
    $f_{d}/f_{s} \times \frac{\B(\Dm\to\Kp\pim\pim)}{\B(\Dsp\to\Kp\Km\pip)}$ & $4.7$ & $4.7$ & $4.7$\Tstrut\\
    Fit model  & $3.4$ & $2.2$ & $2.2$ \\
    Comb. background  & $1.2$ & $1.9$ & $1.5$ \\
    $\DsJp$ background & $0.4$ & $5.0$ & $1.9$ \\
    Simulation statistics  & $1.9$ & $2.1$ & $1.9$ \\
    PID efficiency  & $1.4$ & $1.8$ & $1.5$ \\
    Trigger efficiency & $1.5$ & $1.5$ & $1.5$ \\
    \hline
    Total & $6.6$ & $8.1$ & $6.4$\Tstrut\\
    \hline
    \end{tabular}
%     \caption{\label{tab:totalsystematic}
%     \small %captions should be a little bit smaller than main text
%     Sources of systematic uncertainty for the \BsDsDs branching ratios.}  
\end{center}
\end{table}

\section{Summary and discussion}
\label{sec:Summary}

Inserting the measured yields and relative efficiencies into Eq.~\ref{eq:BR}, along with the $f_{s}/f_{d}$ and 
$\B(\Dm\to\Kp\pim\pim)/\B(\Dsp\to\Kp\Km\pip)$ values taken from~\cite{LHCb-CONF-2013-011}, gives
\begin{align*}
    \frac{\BF(B_{s}^{0}\rightarrow~D_{s}^{(\ast)+}D_{s}^{(\ast)-})}{\BF(\Bd\to\Dsp\Dm)} &=
    4.24 \pm 0.14 \stat \pm 0.27 \syst, \\
    \\
    \frac{\BF(\Bs\to\Dsspm\Dsmp)}{\BF(\Bd\to\Dsp\Dm)} &=
    1.88 \pm 0.08 \stat \pm 0.12 \syst,\\
    \\
    \frac{\BF(\Bs\to\Dssp\Dssm)}{\BF(\Bd\to\Dsp\Dm)} &=
    1.76 \pm 0.11 \stat \pm 0.14 \syst.
\end{align*}

Using the current world average measurement of the $\Bd\to\Dsp\Dm$ branching fraction
of $(7.2 \pm 0.8)\times 10^{-3}$~\cite{PDG2014}, gives
\begin{align*}
    \BF(B_{s}^{0}\rightarrow~D_{s}^{(\ast)+}D_{s}^{(\ast)-}) &= (3.05 \pm 0.10 \pm 0.20 \pm 0.34)\%, \\
    \BF(\Bs\to\Dsspm\Dsmp) &= (1.35 \pm 0.06 \pm 0.09 \pm 0.15)\%, \\
    \BF(\Bs\to\Dssp\Dssm) &= (1.27 \pm 0.08 \pm 0.10 \pm 0.14)\%,
\end{align*}

\noindent where the uncertainties are statistical, systematic, and due to the branching fraction of the normalisation channel, respectively.
%\vspace{6mm}

\begin{figure}[t]
  \begin{center}
    \includegraphics[width=0.85\linewidth]{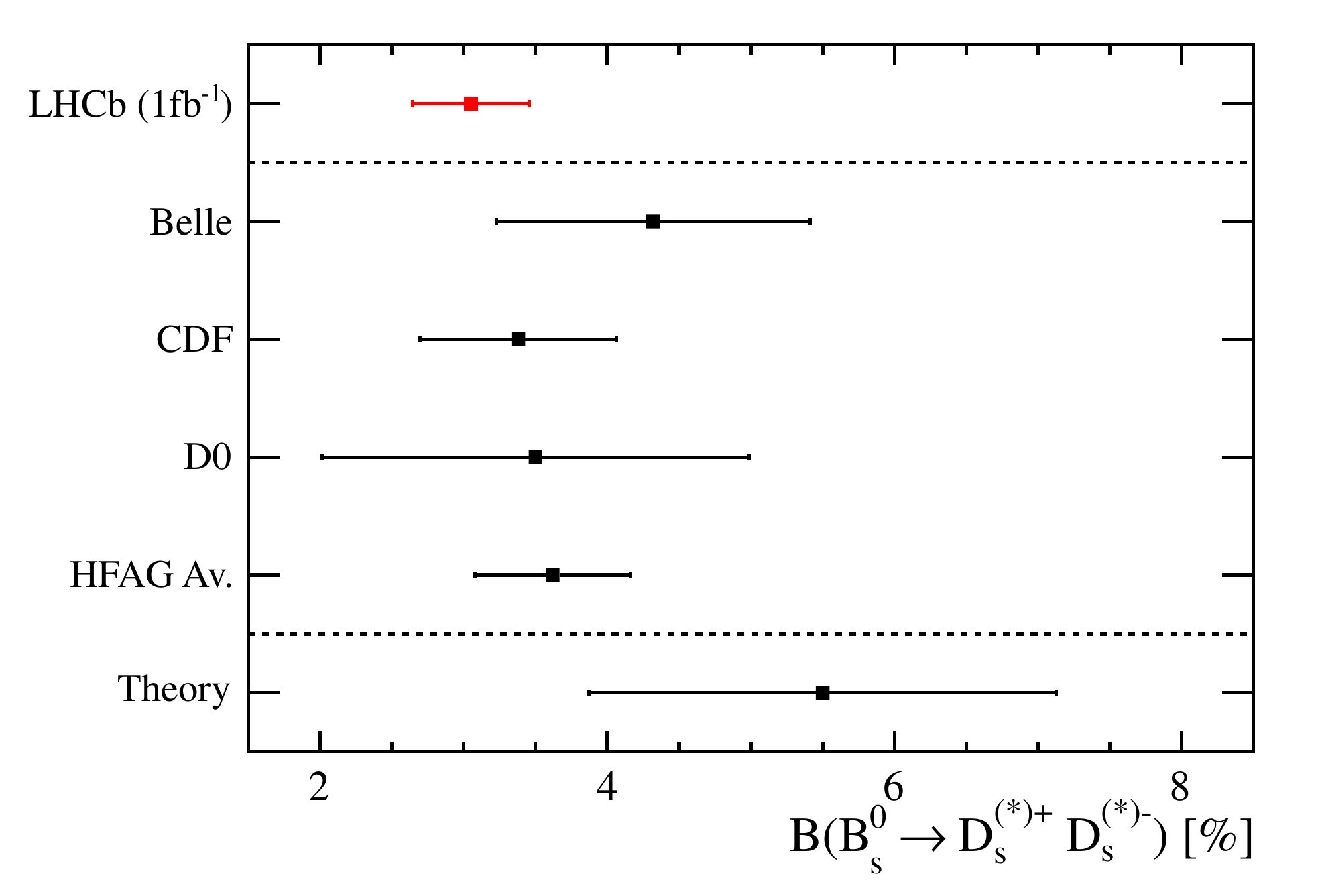}
    \caption{
    \small %captions should be a little bit smaller than main text
    \label{fig:BRsByExpt}The \BsDsDs branching fraction measurements by \belle, \cdf, and \dzero, the current world average, 
    the theoretical prediction from Ref.~\cite{Chua:2011er}, and the new \lhcb result.}
  \end{center}
\end{figure}

Figure~\ref{fig:BRsByExpt} shows the \lhcb measurement of the total \BsDsDs branching fraction,
along with the previous measurements by \belle~\cite{Esen:2012yz}, \cdf~\cite{Aaltonen:2012mg}, and \dzero~\cite{Abazov:2008ig}, the 
average of these previous measurements as calculated by HFAG~\cite{HFAG}, and the theoretical value from Ref.~\cite{Chua:2011er}. 
The theoretical prediction is for a decay time $t=0$, while the measurements integrate over the \Bs meson lifetime; the correction factor for mixing
is known~\cite{DeBruyn:2012wj}, but has not been applied.
The \lhcb result is consistent with
all previous measurements and calculations, and is the most precise determination to date. In addition, the \lhcb measurements
of the individual $\Bs\to\Dsspm\Dsmp$ and $\Bs\to\Dssp\Dssm$ branching fractions are consistent with, and more precise than, all previous
measurements.

Using this measurement of the branching
fraction of \BsDsDs decays to calculate \DGsGs as detailed in Ref.~\cite{Aleksan1993567} gives a value approximately half as large as the
most recent HFAG determination~\cite{HFAG}, suggesting that indeed \BsDsDs decays do not saturate the \CP-even modes~\cite{Lenz:2012mb}. 
The measurements presented in this analysis will be useful in improving the understanding of hadronisation effects in \Bs decays via the
$\decay{\bquark}{\cquark\cquarkbar\squark}$ quark transition, and in determining a precise value of the inclusive branching fraction of 
these $\decay{\bquark}{\cquark\cquarkbar\squark}$ decays.

\section*{Acknowledgements}
 
\noindent We express our gratitude to our colleagues in the CERN
accelerator departments for the excellent performance of the LHC. We
thank the technical and administrative staff at the LHCb
institutes. We acknowledge support from CERN and from the national
agencies: CAPES, CNPq, FAPERJ and FINEP (Brazil); NSFC (China);
CNRS/IN2P3 (France); BMBF, DFG and MPG (Germany); INFN (Italy);
FOM and NWO (The Netherlands); MNiSW and NCN (Poland); MEN/IFA (Romania);
MinES and FANO (Russia); MinECo (Spain); SNSF and SER (Switzerland);
NASU (Ukraine); STFC (United Kingdom); NSF (USA).
We acknowledge the computing resources that are provided by CERN, IN2P3 (France), KIT and DESY (Germany), INFN (Italy), SURF (The Netherlands), PIC (Spain), GridPP (United Kingdom), RRCKI and Yandex LLC (Russia), CSCS (Switzerland), IFIN-HH (Romania), CBPF (Brazil), PL-GRID (Poland) and OSC (USA). We are indebted to the communities behind the multiple open
source software packages on which we depend.
Individual groups or members have received support from AvH Foundation (Germany),
EPLANET, Marie Sk\l{}odowska-Curie Actions and ERC (European Union),
Conseil G\'{e}n\'{e}ral de Haute-Savoie, Labex ENIGMASS and OCEVU,
R\'{e}gion Auvergne (France), RFBR and Yandex LLC (Russia), GVA, XuntaGal and GENCAT (Spain), The Royal Society, Royal Commission for the Exhibition of 1851 and the Leverhulme Trust (United Kingdom).

\addcontentsline{toc}{section}{References}
\setboolean{inbibliography}{true}
%\bibliographystyle{LHCb}
%\bibliography{main,LHCb-PAPER,LHCb-CONF,LHCb-DP,LHCb-TDR}
\ifx\mcitethebibliography\mciteundefinedmacro
\PackageError{LHCb.bst}{mciteplus.sty has not been loaded}
{This bibstyle requires the use of the mciteplus package.}\fi
\providecommand{\href}[2]{#2}

\newpage

%%%%%%%%%%%%%%%%%%%%%%%%%%%%%%%%%%%%%%%%%%
\centerline{\large\bf LHCb collaboration}
\begin{flushleft}
\small
R.~Aaij$^{39}$, 
C.~Abell\'{a}n~Beteta$^{41}$, 
B.~Adeva$^{38}$, 
M.~Adinolfi$^{47}$, 
A.~Affolder$^{53}$, 
Z.~Ajaltouni$^{5}$, 
S.~Akar$^{6}$, 
J.~Albrecht$^{10}$, 
F.~Alessio$^{39}$, 
M.~Alexander$^{52}$
% ...

{\footnotesize \it
% ...
$ ^{5}$Clermont Universit\'{e}, Universit\'{e} Blaise Pascal, CNRS/IN2P3, LPC, Clermont-Ferrand, France\\
$ ^{6}$CPPM, Aix-Marseille Universit\'{e}, CNRS/IN2P3, Marseille, France\\
% ...
$ ^{10}$Fakult\"{a}t Physik, Technische Universit\"{a}t Dortmund, Dortmund, Germany\\
% ...
$ ^{38}$Universidad de Santiago de Compostela, Santiago de Compostela, Spain\\
$ ^{39}$European Organization for Nuclear Research (CERN), Geneva, Switzerland\\
% ...
$ ^{41}$Physik-Institut, Universit\"{a}t Z\"{u}rich, Z\"{u}rich, Switzerland\\
% ...
$ ^{47}$H.H. Wills Physics Laboratory, University of Bristol, Bristol, United Kingdom\\
% ...
$ ^{52}$School of Physics and Astronomy, University of Glasgow, Glasgow, United Kingdom\\
$ ^{53}$Oliver Lodge Laboratory, University of Liverpool, Liverpool, United Kingdom\\
% ...
}
\end{flushleft}
%%%%%%%%%%%%%%%%%%%%%%%%%%%%%%%%%%%%%%%%%%

\end{document}